\documentclass{article}
\usepackage{ijcai24}

\usepackage{times}
\usepackage{soul}
\usepackage{url}
\usepackage[hidelinks]{hyperref}
\usepackage[utf8]{inputenc}
\usepackage[small]{caption}
\usepackage{graphicx}
\usepackage{amsmath}
\usepackage{amsthm}
\usepackage{amssymb}
\usepackage{booktabs}
\usepackage{algorithm}
\usepackage[switch]{lineno}

\usepackage{bm}
\usepackage[noend]{algpseudocode}
\usepackage{xspace}
\usepackage{makecell}
\usepackage{multirow}
\usepackage{array}
\usepackage{color}
\usepackage{colortbl}
\usepackage{enumitem}
\usepackage{pifont}

\newcommand{\eg}{\emph{e.g.,}\xspace}
\newcommand{\ie}{\emph{i.e.,}\xspace}

\newcommand{\baby}{FedPA\xspace}
\newcommand*\samethanks[1][\value{footnote}]{\footnotemark[#1]}

\title{Federated Adaptation for Foundation Model-based Recommendations}

\author{
Chunxu Zhang$^{1,2}$\thanks{This works was done during Chunxu Zhang was an intern at Institute for AI Industry Research, Tsinghua University.}
\and
Guodong Long$^3$\and
Hongkuan Guo$^4$\and
Xiao Fang$^4$\and
Yang Song$^4$\and
Zhaojie Liu$^4$\and
Guorui Zhou$^4$\and
Zijian Zhang$^{1,2}$\and
Yang Liu$^{5}$\thanks{Corresponding authors.}\and
Bo Yang$^{1,2}$\samethanks
\affiliations
$^1$Key Laboratory of Symbolic Computation and Knowledge Engineering of Ministry of Education, China\\
$^2$College of Computer Science and Technology, Jilin University, China\\
$^3$Australian Artificial Intelligence Institute, FEIT, University of Technology Sydney\\
$^4$Kuaishou Technology\\
$^5$Institute for AI Industry Research, Tsinghua University\\
\emails
\{cxzhang19, zhangzj2114\}@mails.jlu.edu.cn,
guodong.long@uts.edu.au,
\{guohongkuan, zhouguorui\}@kuaishou.com,
\{ustcfx0727, liuzj03\}@gmail.com,
ys@sonyis.me,
liuy03@air.tsinghua.edu.cn,
ybo@jlu.edu.cn
}

\begin{document}

\maketitle

\begin{abstract}
    With the recent success of large language models, particularly foundation models with generalization abilities, applying foundation models for recommendations becomes a new paradigm to improve existing recommendation systems. It becomes a new open challenge to enable the foundation model to capture user preference changes in a timely manner with reasonable communication and computation costs while preserving privacy. This paper proposes a novel federated adaptation mechanism to enhance the foundation model-based recommendation system in a privacy-preserving manner. Specifically, each client will learn a lightweight personalized adapter using its private data. The adapter then collaborates with pre-trained foundation models to provide recommendation service efficiently with fine-grained manners. Importantly, users' private behavioral data remains secure as it is not shared with the server. This data localization-based privacy preservation is embodied via the federated learning framework. The model can ensure that shared knowledge is incorporated into all adapters while simultaneously preserving each user's personal preferences. Experimental results on four benchmark datasets demonstrate our method's superior performance. Implementation code is available to ease reproducibility\footnote{\url{https://github.com/Zhangcx19/IJCAI-24-FedPA}}.
\end{abstract}

\section{Introduction}
Recently, the Foundation Models (FMs)~\cite{radford2019language,bommasani2021opportunities,OpenAI2023} emerge rapidly and have made breakthroughs in various AI applications, ranging from language~\cite{alayrac2022flamingo}, vision~\cite{saharia2022photorealistic}, reasoning~\cite{kojima2022large} and recommendation~\cite{geng2022recommendation}. FMs are typically trained on extensive data sources, allowing them to capture and utilize inherent common knowledge. This capability empowers FMs to achieve outstanding performance in various downstream tasks. Applying foundation models to recommendation systems is considered a highly promising direction, which has significantly propelled the state-of-the-art in recommendation system studies~\cite{harte2023leveraging,liu2023pre,lin2023can}.

Two new open challenges have been encountered when we introduce the foundation models into a practical recommendation system. First, given the fast changes in user preference, how to timely update the foundation models-based recommendation system with reasonable cost on communication and computation. An on-device parameter-efficient fine-tuning mechanism is desired to be incorporated with the foundation models. The second challenge is how to tackle the privacy-sensitive data of users that is needed to train or fine-tune the foundation model in a recommendation system. 

Federated learning trains the global model by iterative model parameters transmission between server and clients without accessing private client data~\cite{mcmahan2017communication,miao2023task,zhong2023personalized,liu2024icde}. Due to its excellent privacy-preserving properties, federated learning has become a popular privacy protection enhanced scheme for recommendation, named federated recommendation systems~\cite{chai2020secure,yang2020federated,wu2022federated,sun2022survey,zhang2023dual,li2023federated}. Trained on extensive data, the foundation model's attribute embedding and prediction function retain valuable general domain knowledge and user decision logic. Integrating federated learning into foundation model-based recommendation systems can benefit from this shared knowledge while ensuring privacy preservation. However, effectively addressing this integration remains unsettled.

Several challenges must be addressed urgently to adapt foundation model-based recommendations with federated learning frameworks. \textit{\textbf{Challenge 1: efficient user personalization modeling.}} Given the substantial preference differences across isolated clients, the uppermost goal for this recommendation system is effective personalization modeling of diverse users under the privacy protection limitation. At the same time, there are usually concerns about efficiency when using large-scale foundation models, which is more urgent in federated learning that requires end-device optimization. \textit{\textbf{Challenge 2: common knowledge and user personalization fusion.}} Common knowledge maintained in the pre-trained foundation model can incorporate insights from collective user behavior. By integrating it with user personalization, the federated recommendation system can leverage the collective intelligence of the user community, leading to improved decision-making. However, the importance of common knowledge and user personalization varies across users, and ineffective knowledge fusion can lead to information confusion and misleading recommendations. Hence, balancing general knowledge and individual user preference emerges as a main challenge.

In this paper, we present a novel method, named \textbf{Fed}erated recommendation with \textbf{P}ersonalized \textbf{A}dapter (\textbf{\baby}), to explore the federated foundation model for recommendation. Our method leverages the pre-trained model as the foundation, allowing us to incorporate common knowledge and optimize the federated recommendation system from a well-established starting point. To capture user personalization efficiently, we propose a personalized adapter to deploy on the client, which can learn individual user preferences in a lightweight manner.
Then, we learn these adapters with the pre-trained model in an adaptive fusion manner to balance the collaboration of common knowledge and user personalization in federated optimization. 

Motivated by research about neural network optimization~\cite{li2018measuring,aghajanyan2020intrinsic}, the model parameter solution for the target task usually resides within an intrinsic dimension. In the federated recommendation system, each user has task-specific intrinsic parameter space that can be learned from personal data. To this end, we design a low-rank adapter to learn user personalization in a lightweight manner. Particularly, we develop two levels of personalization to accommodate the recommendation scenario, including user-level and user-group-level. They cater to the unique preferences of individual users while also capturing and leveraging shared patterns and preferences within specific user groups. Furthermore, we design an adaptive gate learning mechanism that dynamically learns the weights for common knowledge and user personalization, enabling effective knowledge fusion. During federated optimization, our \baby focuses on updating only the parameters relevant to user-specific modeling and the others are frozen and exempted from optimization, leading to a significant reduction in communication cost and achieving faster convergence.

We assess the performance of \baby on four benchmark datasets and compare it with various advanced baselines. Experimental results consistently demonstrate that our method outperforms the baselines by a significant margin. In addition, we conduct comprehensive experiments to analyze \baby's ability to capture user personalization and the impact of common knowledge in the pre-trained model on federated recommendation systems. Furthermore, we validate the model's feasibility in real-world applications. By distilling the pre-trained model into a smaller size, we address the computational and storage challenges of deploying the pre-trained model on edge devices. Additionally, we enhance privacy protection by leveraging the Local Differential Privacy technique. Experimental results demonstrate \baby's stable performance with distilled smaller models and privacy preservation, affirming its practical applicability. To summarize, the \textbf{main contributions} are listed as follows,
\begin{itemize}
    \item For the first time, we investigate the federated adaption paradigm for foundation model-based recommendation, named \baby. It enables the integration of the rich knowledge encapsulated within pre-trained models while upholding privacy protection for users. 
    \item We present a personalized low-rank adapter to learn user personalization from user-level and user-group-level in a lightweight manner. Furthermore, we design an adaptive gate learning mechanism to dynamically learn weights, allowing for the effective fusion of common knowledge with user personalization.
    \item Extensive experiments on four benchmark datasets demonstrate the superior performance of \baby against advanced baselines. Additionally, \baby also shows excellent feasibility in deploying on clients with limited computation capability and strengthening user privacy protection in federated recommendation systems.
\end{itemize}

\section{Related Work}
\subsection{Foundation Models for Recommendation}
Pre-training in Natural Language Processing (NLP)~\cite{qiu2020pre} has witnessed significant progress, with language models like GPT~\cite{brown2020language} and BERT~\cite{devlin2018bert} achieving state-of-the-art results. The pre-training and fine-tuning paradigm allows for the extraction of valuable knowledge and eliminates the need to train new models from scratch. Given its remarkable benefits, increasing research has been on developing foundation models for recommendation systems~\cite {liu2023pre,wu2023survey}. The essential learning objective of recommendation is to estimate the user's preferences for a certain item set. By incorporating the foundation models into the recommendation system, it can absorb valuable knowledge and improve the system's ability for characteristics extraction and user decision pattern learning. However, existing foundation models for recommendation rely on collecting personal data from users to optimize, which poses severe risks to protecting user privacy.

\begin{figure*}[!t]
    \centering
    \setlength{\abovecaptionskip}{-0.1mm}
    \setlength{\belowcaptionskip}{-3mm}
    \includegraphics[width=0.75\linewidth]{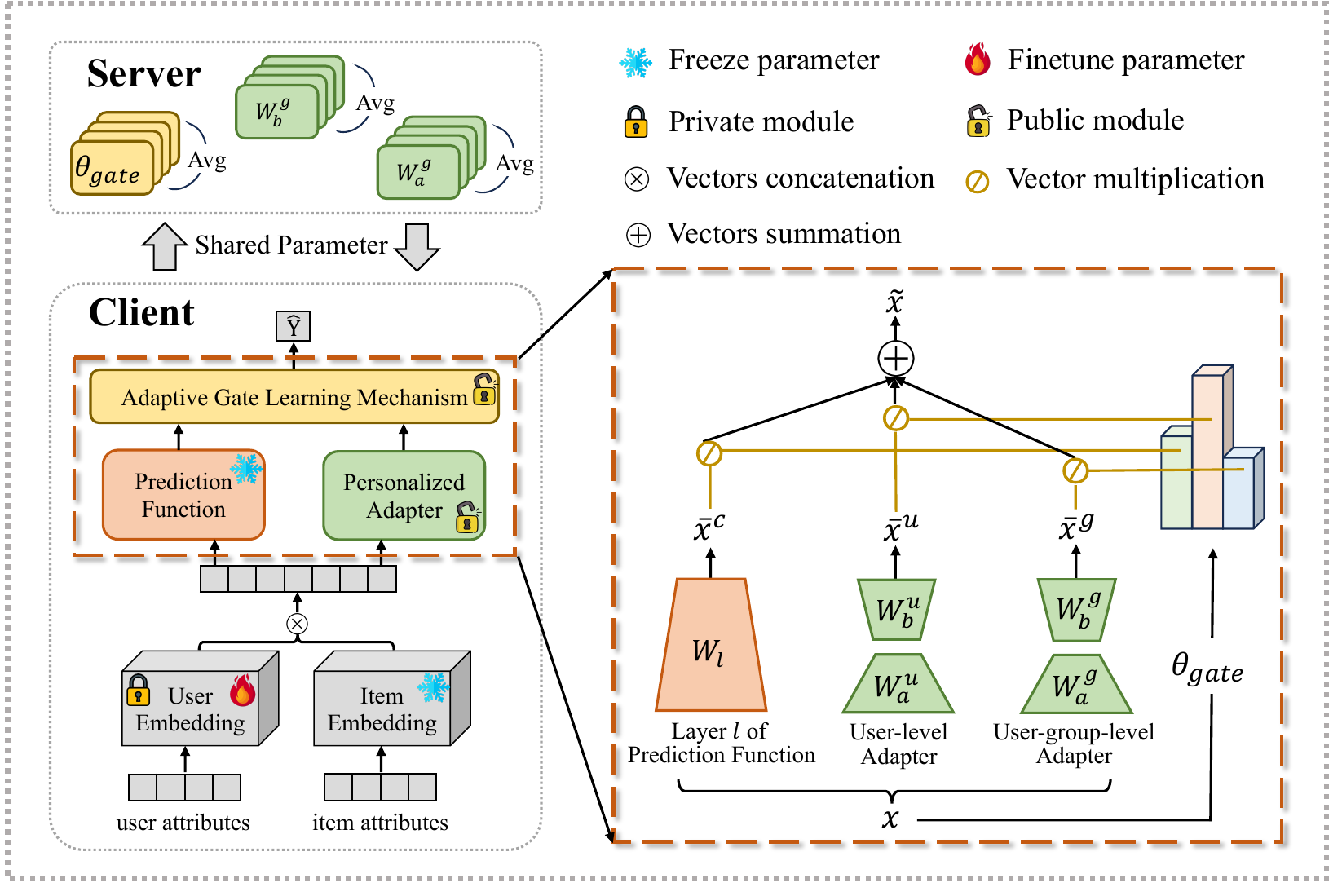}
    \caption{The framework of \baby. The left part represents the workflow of our method. Each client learns the local recommendation model based on personal data, initializing it with parameters from the pre-trained model. During training, we only update the parameters about user personalization modeling and keep others frozen. The server is responsible for globally aggregating the shared parameters to transmit common information among clients. The right part illustrates the details of employing the adaptive gate learning mechanism to fuse the common information from each layer of the prediction function and user personalization from the personalized adapter at two granularities. It is worth noting that the user-level adapter is a private module and the user-group-level adapter is a public module.}
    \label{fig:framework}
\end{figure*}

\subsection{Federated Recommendation System}
Federated recommendation system is a rising service paradigm that can learn the model in a privacy-preserving manner~\cite{lin2020meta,perifanis2022federated,qu2023semi,zhang2023ifedrec}. Existing studies focus on developing federated recommendation models with mainstream recommendation architectures, \eg matrix factorization~\cite{chai2020secure} and neural collaborative filtering~\cite{perifanis2022federated}, popular recommendation tasks, \eg POI prediction~\cite{zhang2023fine} and multi-domain recommendation~\cite{liu2023federated}, and user privacy protection enhancement~\cite{liu2023privaterec,huang2023randomization}. In this paper, we investigate the integration of foundation model into federated recommendation system, which can utilize the inherent common knowledge in the pre-trained model and prosper the powerful federated recommendation system.

\section{Methodology}
We present a novel federated adaption paradigm for foundation model-based recommendation, named \textbf{Fed}erated recommendation with \textbf{P}ersonalized \textbf{A}daptor (\textbf{\baby}). In this section, we first provide an overview of the framework architecture. Then, we delve into the components to elucidate the details and summarize the workflow into an optimization algorithm. Following that, we engage in a discussion on the feasibility of deploying our \baby in physical applications. Finally, we develop a privacy-protection enhanced \baby that can strengthen the system's protection for user privacy.

\subsection{Framework Overview}
The pre-trained recommendation model is learned from a large amount of publicly available data, embodying rich knowledge. It can effectively characterize user and item attributes and possesses strong predictive capabilities for user decision patterns. Effective utilization of this common knowledge contributes to building a more powerful federated recommendation model. To achieve this, we take the pre-trained model as the foundation model for the federated recommendation system, enabling federated optimization from a favorable starting point.

To efficiently capture user-specific preferences, we propose a low-rank adapter that models user personalization from both the user-level and user-group-level perspectives. Additionally, we design an adaptive gate learning mechanism that effectively integrates common knowledge and personalized knowledge for better user modeling. The model architecture is illustrated in Figure 1. In this paper, we utilize an existing recommendation architecture as the base model, and it can be easily extended to other popular architectures.

\subsection{Base Model}
Given user attributes, item attributes, and user interaction records, we adopt a widely used two-tower recommendation model architecture. Specifically, the model consists of two input branches, one for learning user representations based on user attributes and the other for learning item representations based on item attributes. These representations are then fed into a prediction function to estimate user preferences for items. The user interaction records serve as supervision information to guide the model in updating its parameters.

\noindent \textbf{User Embedding Module.}
There are certain attribute information available for the users, \eg user active degree. For each attribute $i$, we construct a learnable embedding table $E_i \in \mathbb{R}^{p \times d}$, where $p$ is the total attribute categories and $d$ is the embedding dimension. Then, for each user $u$ with attributes $A_u$, we retrieve the embedding vectors from all the attribute embedding tables based on attribute values and obtain the user representation $u_r$ by concatenating them,
\begin{equation}
    u_r = {\rm Concat}({E_i(A_u^i)}_{i=1}^{|A_u|})
\end{equation}
where $|A_u|$ is the total attribute number of user $u$.

\noindent \textbf{Item Embedding Module.}
For each item $v$, we adopt a similar approach as the user embedding module to construct the item embedding tables and obtain item representation $v_r$,
\begin{equation}
    v_r = {\rm Concat}({E_i(A_v^i)}_{i=1}^{|A_v|})
\end{equation}
and $|A_v|$ is the total attribute number of item $v$.

\noindent \textbf{Prediction Function Module.}
Given user representation $u_r$ and item representation $v_r$, we use a simple MLP (Multi-Layer Perceptron) as the prediction function to estimate user preferences for items,
\begin{equation}
    \hat{Y}_{uv} = {\rm MLP}({\rm Concat}(u_r, v_r))
\end{equation}

\noindent \textbf{Loss Function.}
To update the model parameters, we construct a loss function that encourages the model's predictions to be as close as possible to the true labels. For common implicit feedback recommendation tasks, where the label value is $Y_{uv}=1$ when a user $u$ interacts with an item $v$ and $Y_{uv}=0$ otherwise, we use binary cross-entropy as the loss function,
\begin{equation}
    L(\theta_{base}) = -\frac{1}{|D|} \sum_{Y_{u,v} \in D} Y_{uv} \log \hat Y_{u,v} + (1-Y_{uv}) \log (1 - \hat Y_{u,v})
\end{equation}
where $\theta_{base}$ is the model parameter, including user embedding tables $\theta_{ue}$, item embedding tables $\theta_{ie}$ and the MLP parameters $\theta_{mlp}$. $D$ is the user-item interaction record set, and $|D|$ is the total interactions number.

\subsection{Personalized Adapter}
Existing federated recommendation methods typically model user personalization by preserving partial model parameters locally. However, it limits direct access from other clients and potentially hinders the collaborative context utilization. Moreover, the personal data on each client is generally limited, which can introduce biases and compromise model performance. To overcome the challenge, we propose to learn user personalization from two perspectives, \ie user-level and user-group-level. Particularly, we devise a personalized adapter applied to the prediction function module due to its crucial role in predicting user preferences. 

Drawing inspiration from research on neural network optimization, model parameters can be embedded within an intrinsic dimension~\cite{li2018measuring,aghajanyan2020intrinsic}.  
To this end, we propose a low-rank adapter that leverages low-rank matrices to model user-specific knowledge on each client. This approach offers two prominent advantages: First, it can learn user personalization from user-specific and user groups with similar characteristics, which enhances the system's ability to model individual user preferences. Besides, the low-rank matrices introduce only a small number of parameters, making it a parameter-efficient solution. 

\noindent \textbf{Low-Rank adapter.}
For each layer $l$ of MLP in the prediction function module, it maps the input $x \in \mathbb{R}^{d}$ into a new space $\overline{x} \in \mathbb{R}^{k}$ with the weight matrix $W_l \in \mathbb{R}^{k \times d}$. We intensify the personalization learning by adding a low-rank decomposition matrix $W_{lr}=W_a W_b$, where $W_a \in \mathbb{R}^{k \times r}$ and $W_b \in \mathbb{R}^{r \times d}$, and the rank $r \ll min(d,k)$. Then, the forward pass of each layer can be modified as follows,
\begin{equation}
    \overline{x} = W_l x + W_a W_b x
\end{equation}
where $W_{lr}$ is responsible for learning user personalization. In the context of recommendation tasks, we develop two levels of personalization: user-level personalization and user-group-level personalization. They cater to individual user preferences as well as capturing patterns and preferences shared within specific user groups. 

\noindent \textbf{User-Level Personalization.}
For the user-level personalization, we aim to learn a specific low-rank adapter for each user so the parameters $W_a^u$ and $W_b^u$ would be preserved locally and not be shared globally. For the user $u$, we formulate the user-specific low-rank adapter as follows,
\begin{equation}
    \overline{x}^u = W_a^u W_b^u x
\end{equation}

\noindent \textbf{User-Group-Level Personalization.}
For the user-group-level personalization, we aim to learn the same low-rank adapter for users in a specific group. In recommendation systems, users with similar characteristics tend to share similar preferences. To fully leverage this information, we learn multiple groups of low-rank adapters. For each user group $g$, we formulate the group-specific low-rank adapter as follows,
\begin{equation}
    \overline{x}^g = W_a^g W_b^g x
\end{equation}
Users within the same group share parameter $W_a^g$ and $W_b^g$. It is worth noting that users in the system can be grouped in multiple ways, meaning each user $u$ can belong to multiple groups $\{g_i\}_{i=1}^{total}$. For example, a user can belong to the ``young adults" group while also to the ``highly active degree" group. By categorizing users from multiple orthometric perspectives, our model can learn more detailed personalized parameters and enhance the user preference capture.

\subsection{Adaptive Gate Learning Mechanism}
By incorporating the low-rank adapter, the prediction function module can learn both the shared decision patterns among users and the personalized decision logic at two granularities, \ie user-level and user-group-level. To effectively combine common knowledge and personalized knowledge, we propose an adaptive gate learning mechanism that dynamically assigns weights to each decision branch. Specifically, we utilize a two-layer non-linear mapping to learn the weights for the branches based on the input. The fusion process can be formulated as follows,
\begin{equation}
\begin{split}
    \widetilde{x} = & {\rm Sum}(softmax(W_2 Relu(W_1 x)) \odot \\  
    & [\overline{x}^c, \overline{x}^u, \{\overline{x}^{g_i}\}_{i=1}^{total}])
\end{split}
\end{equation}
where $W_1$ and $W_2$ are the parameters of the adaptive gate learning mechanism, $\overline{x}^c$ is the output of the MLP layer in the prediction function module. $\odot$ and ${\rm Sum}(\cdot)$ denote element-wise multiplication and summation calculation, respectively. The parameters of the adaptive gate learning mechanism can be updated based on gradients along with other model parameters, eliminating the cumbersome manual setting of hyperparameters. This approach enhances the flexibility of multi-branch fusion, allowing for adaptive adjustment of the weights for all branches during the model's training stages.

\subsection{Optimization and Model Scalability}
\subsubsection{Optimization Objective}
In the federated recommendation system, each user $u$ acts as a client and trains the local recommendation model according to the personal dataset $D_u$. Let $L_u(\theta^u_{all})$ denote the local model's loss function, where $\theta^u_{all}$ contains of base model parameter $\theta_{base}$, low-rank adapter parameter $\theta_{lr}^u$ and $\{\theta_{lr}^{g_i}\}_{i=1}^{total}$ and adaptive gate learning mechanism parameter $\theta_{gate}$. The overall optimization of the federated recommendation model can be formulated as follows,
\begin{equation}
    \min_{\{\theta_{all}^1,...,\theta_{all}^n\}} \sum_{u=1}^n \frac{1}{n} \mathcal{L}_u(\theta_{all}^u)
\end{equation}
where $n$ is the total number of clients in the federated recommendation system. Here, we employ a naive average aggregation approach to optimize system parameters. It is also possible to use a more flexible weighted aggregation approach further to enhance the optimization objective~\cite{mcmahan2017communication,wang2020tackling}.

\subsubsection{Efficient Parameter Update}
In our method, we employ the pre-trained model as the foundation model to warm-start the federated recommendation system. It enables us to leverage the common knowledge inherent in the pre-trained model, thereby enhancing the system's learning capability. Since the pre-trained model is trained on large-scale data, the pre-trained model has already gained an in-depth understanding of the item characteristics and general decision patterns. To leverage this knowledge effectively, we propose to freeze the item embedding module and prediction function module during federated optimization. However, since the federated recommendation system operates on different user sets, we continue to update the user embedding module to adapt to the specific characteristics. To summarize, during the federated optimization process, we focus solely on updating the parameters related to personalized user modeling, \eg user embedding, low-rank adapter, and adaptive gate learning mechanism parameters. It effectively saves computational and communication costs.

\subsubsection{Discussion about Model Scalability}
The pre-trained recommendation models learned with abundant computational resources often have complex structures and large sizes in real-world scenarios. Deploying such models directly to the clients poses significant challenges in terms of limited storage and computational capability. To address this issue, we discuss the feasible solution for applying the proposed framework in practical scenarios. Specifically, we leverage the knowledge distillation (KD) technique~\cite{gou2021knowledge} to distill the pre-trained model into a smaller-sized recommendation model that can be accommodated by client devices. We then utilize this distilled model to warm up the federated recommendation system. This approach effectively enhances the scalability of the proposed framework in real-world applications.

\subsection{Privacy-Preserving Enhanced \baby}
The distributed training nature of federated learning avoids direct exposure to private user data. To further mitigate the risk of the server inferring user privacy through model parameter reverse engineering, we integrate the Local Differential Privacy technique~\cite{choi2018guaranteeing} into our method, whose basic insight is to add a zero-mean Laplacian noise to the shared model parameters before uploaded to the server. Our method's shared model parameters include user group-level low-rank adapter and the adaptive gate learning mechanism parameters. By adjusting the intensity of noise, we can control the privacy protection capability of the system, that is, increasing the intensity of the noise enhances the effectiveness of privacy protection.

\begin{table*}[!t]
\renewcommand{\arraystretch}{1.2}
\setlength\tabcolsep{0.6pt}
\centering
\setlength{\abovecaptionskip}{1mm}
\setlength{\belowcaptionskip}{-3mm}
\small
\begin{tabular}{p{50pt}<{\centering}|p{80pt}||p{40pt}<{\centering}p{40pt}<{\centering}|p{40pt}<{\centering}p{40pt}<{\centering}|p{40pt}<{\centering}p{40pt}<{\centering}|p{40pt}<{\centering}p{40pt}<{\centering}}
\hline 
\multirow{2}{*}{\textbf{Method}} & \textbf{Dataset} & \multicolumn{2}{c|}{\textbf{KuaiRand-Pure}} &
\multicolumn{2}{c|}{\textbf{KuaiRand-small}} &
\multicolumn{2}{c|}{\textbf{KuaiSAR-S}} &
\multicolumn{2}{c}{\textbf{KuaiSAR-R}} \\
\cline{2-10}
& \textbf{Metric} & AUC & Precision & AUC & Precision & AUC & Precision & AUC & Precision \\
\hline
\multirow{5}{*}{\textbf{w/o Warm}} & FedNCF & 68.17 & 73.33 & 69.35 & 65.69 & 55.19 & \textbf{81.82} & 70.83 & 65.45 \\
& PFedNCF & 62.99 & 72.09 & 63.73 & 62.23 & 55.31 & 79.44 & 67.69 & 63.81 \\
& FedRecon & 65.21 & 61.72 & 68.81 & 65.32 & \textbf{58.56} & 76.57 & 66.65 & 62.68 \\
& PFedRec & 59.48 & 70.80 & 61.09 & 61.33 & 56.71 & 79.42 & 68.32 & 63.83 \\
\cline{2-10}
& \textbf{\baby w/o Warm} & \textbf{68.44*} & \textbf{73.75*} & \textbf{69.65*} & \textbf{66.69*} & 57.58 & 81.04 & \textbf{71.30*} & \textbf{65.61*} \\
\hline
\hline
\multirow{5}{*}{\textbf{w/ Warm}} & Warm\_FedNCF & 69.90 & 74.15 & 70.54 & 65.83 & 59.95 & 79.03 & 71.47 & 66.38 \\
& Warm\_PFedNCF & 62.78 & 71.75 & 62.58 & 61.59 & 57.86 & 79.24 & 66.82 & 63.43 \\
& Warm\_FedRecon & 70.02 & 74.22 & 70.75 & 66.56 & 55.68 & 78.54 & 65.49 & 61.31 \\
& Warm\_PFedRec & 62.71 & 71.73 & 63.96 & 61.83 & 59.74 & 79.30 & 67.50 & 63.81 \\
\cline{2-10}
& \textbf{\baby} & \textbf{70.28*} & \textbf{75.12*} & \textbf{71.14*} & \textbf{66.86} & \textbf{61.99*} & \textbf{86.98*} & \textbf{72.21*} & \textbf{66.70*} \\
\hline
\end{tabular}
\caption{Experimental results of baselines and our method on four datasets. ``\textbf{w/o Warm}" (``\textbf{w/ Warm}") denotes training the federated recommendation system without (with) a pre-trained model. The best results are bold. ``\textbf{{\Large *}}'' indicates the statistically significant improvements (i.e., two-sided t-test with $p<0.05$) over the best baseline.}
\label{main_results}
\end{table*}

\section{Experiment}
In this section, we conduct comprehensive experiments to verify the efficacy and illustrate a deep analysis of various aspects of our proposed \baby.
\subsection{Experimental Setup}
\subsubsection{Datasets}
We evaluate \baby on four practical industrial recommendation datasets collected from the short video platform Kuaishou \footnote{https://www.kuaishou.com/cn}, \ie KuaiRand \footnote{https://kuairand.com/} (KuaiRand-Pure and KuaiRand-small) and KuaiSAR \footnote{https://kuaisar.github.io/} (KuaiSAR-S and KuaiSAR-R). For dataset split, we first divide each dataset into two subsets: one for training the foundation model and the other for training the federated recommendation system. The dataset for the federated recommendation system is further split into train, validation, and test sets for each user based on interaction timestamps, with a ratio of 6:2:2.

\subsubsection{Evaluation Protocols}
We evaluate the model performance by calculating the evaluation metrics on the test set of private users. Specifically, we take widely used AUC (Area Under Curve) and Precision as two evaluation metrics. All experimental results are in units of 1e-2 and the average values of five individual runs. For the user group level low-rank personalization, we group users based on their attribute values. It is important to note that the clients can update the corresponding user group low-rank adaptor parameters based on their own attribute values locally without exposing the attribute values to the server. Additionally, we have incorporated Local Differential Privacy technique to further protect user privacy, hence user privacy protection can be ensured in our method.

\subsubsection{Baselines}
This paper concentrates on developing a personalized federated recommendation system and investigating the assistance provided by the common knowledge contained in pre-trained models. To assess the feasibility and effectiveness of our proposed \baby, we compare it with two types of baseline models: (1) Train and evaluate the model on the private user dataset without warm-staring from the pre-trained model, \ie \textbf{w/o Warm}. (2) Warm-start the federated recommendation models with the pre-trained model, \ie \textbf{w/ Warm}. Specifically, we select four representative personalized federated recommendation models, including FedNCF~\cite{perifanis2022federated}, PFedNCF~\cite{pillutla2022federated}, FedRecon~\cite{singhal2021federated} and PFedRec~\cite{zhang2023dual}, and the corresponding warm-starting variants as baselines. Besides, we remove the warm-starting strategy from our method, named \textbf{\baby w/o Warm}, to assess the contribution of our personalization modeling insight for the federated recommendation system.

\subsection{Comparison with Baselines}
Table~\ref{main_results} shows the performance of AUC and Precision on four datasets. We provide a summary of the experimental results and discuss noteworthy observations as follows, 

(1) \textbf{Compared to the naive FedNCF, existing user personalization modeling techniques provide modest performance improvements and, in some cases, can even degrade the model performance.} Existing federated recommendation models focus on modeling user preferences based on user (item) ID attributes. The key idea behind these personalization techniques is to learn user-specific model parameters (\eg prediction function module) via personal data. However, when the system can leverage additional attributes, it provides mode auxiliary information for model optimization and increases learning difficulty. Consequently, selecting specific model parameters exclusively for modeling user personalization loses effectiveness. In contrast, our method introduces low-rank matrices to modeling personalization from both user-level and user-group-level, which ensures common knowledge sharing and enhances model performance.

\begin{table*}[!t]
\renewcommand{\arraystretch}{1.2}
\setlength\tabcolsep{0.6pt}
\centering
\setlength{\abovecaptionskip}{1mm}
\setlength{\belowcaptionskip}{-2mm}
\small
\begin{tabular}{p{40pt}||p{65pt}<{\centering}|p{40pt}<{\centering}|p{70pt}<{\centering}|p{70pt}<{\centering}|p{55pt}<{\centering}|p{85pt}<{\centering}|p{40pt}<{\centering}}
\hline 
\multirow{2}{*}{\textbf{Model}} & \multirow{2}{*}{\textbf{Warm\_FedNCF}} & \multirow{2}{*}{\textbf{w/ UP}} & \multicolumn{4}{c|}{\textbf{w/ GP}} & \multirow{2}{*}{\textbf{\baby}} \\
\cline{4-7}
& & & user active degree & register days range & onehot feat10 & follow user num range \\
\hline
AUC & 69.90 & 70.01 & 70.05 & 70.17 & 70.02 & 70.04 & \textbf{70.28} \\
Precision & 74.15 & 74.26 & 74.18 & 75.00 & 75.04 & 74.98 & \textbf{75.12} \\
\hline
\end{tabular}
\caption{Experimental results for user-level (\textbf{w/ UP}) and user-group-level (\textbf{w/ GP}) low-rank personalization analysis on KuaiRand-Pure. Specifically, we select multiple user attributes, \eg user active degree, to guide the user grouping in user-group-level personalization.}
\label{personalization_results}
\end{table*}
(2) \textbf{Warm-starting the federated recommendation systems with pre-trained models can improve their performances in most cases.} For instance, when incorporating the warm-start strategy, FedNCF achieves 2.54\% and 8.62\% performance gains on KuaiRand-Pure and KuaiSAR-S, respectively. The pre-trained model has already learned general user preferences. By transferring the common knowledge to the federated recommendation system, it is able to supplement the beneficial information and alleviate the performance bottleneck caused by distributed data storage. In our method, we devise an adaptive gate learning mechanism. It can dynamically learn the weights for an effective blend of common knowledge and personalized knowledge and hence achieves state-of-the-art performance.

(3) \textbf{Our \baby is a communication-efficient federated foundation model for recommendation.} In the federated optimization phase, our method only updates the parameters relevant to user personalization modeling, leading to significant savings in communication overhead. For example, on the KuaiRand dataset, the number of trainable parameters of baseline models is 13,649, while only 8,189 for our method, resulting in a 40\% reduction. In federated recommendation systems, the number of clients is typically extensive, and frequent parameter communication between clients and servers presents a substantial challenge to system optimization. Our method effectively reduces communication overhead, making it suitable for deployment in real-world environments.

\subsection{Low-Rank Personalization Analysis}
Our \baby enables the learning of low-rank personalization at both the user and user-group levels, providing comprehensive modeling of user preferences from multiple perspectives. To make an in-depth analysis of the efficacy of the two forms of personalization, we conduct two model variants: one focusing on user-level personalization and the other on user-group-level personalization. Specifically, we take FedNCF as the baseline and incorporate two forms of personalization, denoted as \textbf{w/ UP} (with user-level) and \textbf{w/ GP} (with user-group-level), respectively. Given the user grouping based on their attributes, we conduct the experiments according to multiple user attributes to make a comprehensive analysis. As shown in Table~\ref{personalization_results}, incorporating either user-level or user-group-level low-rank personalization into FedNCF can improve its performance. Therefore, combining these two forms of personalization allows them to complement each other and achieve superior performance.

\begin{table}[!ht]
\renewcommand{\arraystretch}{1.2}
\setlength\tabcolsep{0.6pt}
\centering
\setlength{\abovecaptionskip}{1mm}
\small
\begin{tabular}{p{40pt}||p{40pt}<{\centering}|p{40pt}<{\centering}|p{40pt}<{\centering}|p{40pt}<{\centering}}
\hline 
\textbf{Model} & \textbf{FZ\_UE} & \textbf{FZ\_IE} & \textbf{FZ\_PF} & \textbf{\baby} \\
\hline
AUC & 65.55 & \textbf{70.31} & 70.28 & 70.28 \\
Precision & 72.67 & 75.16 & \textbf{75.27} & 75.12 \\
\hline
\end{tabular}
\caption{Effect of freezing different modules of the pre-trained model on federated optimization on KuaiRand-Pure. ``\textbf{FZ\_UE}", ``\textbf{FZ\_IE}" and ``\textbf{FZ\_PF}" denote freezing user embedding, item embedding and prediction function, respectively.}
\label{knowledge_effect_results}
\end{table}
\subsection{Effect of Common Knowledge on Federated Optimization}
In the recommendation model, different modules have specific roles. User embedding and item embedding focus on attribute information learning, while the prediction function captures user decision patterns. To further explore the impact of the common knowledge inherent in the pre-trained model on federated optimization, we assess the specific effects of each module in the recommendation model on model performance. Specifically, we conduct warm-start experiments using pre-trained models, freezing user embedding, item embedding, and the prediction function separately, and analyze the experimental results.

From Table~\ref{knowledge_effect_results}, we can summarize two conclusions: \textbf{(1) Fixing user embedding during the optimization process of federated recommendation system leads to a significant decline in model performance.} This is because the pre-trained model and the federated recommendation system are trained on different user sets, and the variations between user sets make it challenging for the pre-trained user embedding to adapt well to the federated recommendation system, and consequently resulting in a degradation of performance. \textbf{(2) Updating item embedding or the prediction function in the federated recommendation system leads to further performance improvement.} This finding aligns with our expectations, as updating a larger number of parameters facilitates the model in learning intricate user preferences. Our \baby, which simultaneously fixes item embedding and the prediction function, greatly alleviates communication overhead, presenting an effective compromise to balance the performance and cost of federated recommendation systems.

\subsection{Lightweight \baby with KD}
In the physical setting, the service provider generally learns a large-scale pre-trained model, which poses challenges in terms of computational and storage capabilities when deployed directly to clients. To fill the gap, we develop a lightweight \baby with knowledge distillation technique. Specifically, we first distill a small-scale model from the pre-trained large-scale model and then deploy it on each client as the base model. For a comprehensive investigation, we distill three different sizes, \ie 8-(8, 1), 4-(32, 8, 1) and 4-(8, 1), from the original model whose size is 8-(32, 8, 1) by adjusting the embedding dimension or the prediction function architecture. As shown in Table~\ref{kd_results}, employing the distilled small-scale models for warm-starting the federated recommendation system not only preserves the model's performance but also yields performance improvements. This finding further strengthens our method's viability in real-world applications.
\begin{table}[!t]
\renewcommand{\arraystretch}{1.2}
\setlength\tabcolsep{0.6pt}
\centering
\setlength{\abovecaptionskip}{1mm}
\small
\begin{tabular}{p{40pt}||p{45pt}<{\centering}|p{45pt}<{\centering}|p{45pt}<{\centering}|p{45pt}<{\centering}}
\hline 
\textbf{Model} & \textbf{8-(32, 8, 1)} & \textbf{8-(8, 1)} & \textbf{4-(32, 8, 1)} & \textbf{4-(8, 1)} \\
\hline
AUC & 70.28 & \textbf{71.41} & 71.12 & 70.61 \\
Precision & 75.12 & 75.47 & \textbf{77.49} & 76.81 \\
\hline
\end{tabular}
\caption{Experimental results of warm-starting the federated recommendation system with different-sized models by knowledge distillation on KuaiRand-Pure.}
\label{kd_results}
\end{table}

\subsection{Privacy-Enhanced \baby}
We conduct experiments to evaluate the privacy-protection enhanced \baby by integrating the Local Differential Privacy technique. Particularly, we set the Laplacian noise with different intensities, \eg from 0.1 to 0.5 with an interval of 0.1, to observe the effect, and the results are summarized in Table~\ref{ldp_results}. As the noise intensity increases, the model performance deteriorates, while the decline is slight when the noise is not excessively large. Therefore, a moderate noise intensity, \eg 0.2 is desirable to strike a balance between model performance and system privacy protection capability.
\begin{table}[!ht]
\renewcommand{\arraystretch}{1.2}
\setlength\tabcolsep{0.6pt}
\centering
\setlength{\abovecaptionskip}{1mm}
\small
\begin{tabular}{p{40pt}||p{30pt}<{\centering}|p{30pt}<{\centering}|p{30pt}<{\centering}|p{30pt}<{\centering}|p{30pt}<{\centering}|p{30pt}<{\centering}}
\hline 
\textbf{Intensity} & \textbf{0} & \textbf{0.1} & \textbf{0.2} & \textbf{0.3} & \textbf{0.4} & \textbf{0.5} \\
\hline
AUC & \textbf{70.28} & 70.06 & 69.93 & 69.92 & 69.80 & 69.89 \\
Precision & \textbf{75.12} & 74.17 & 74.35 & 74.28 & 74.20 & 74.16 \\
\hline
\end{tabular}
\caption{Experimental results of privacy-protection enhanced \baby with various noise intensity on KuaiRand-Pure.}
\label{ldp_results}
\end{table}

\section{Conclusion}
In this paper, we develop \baby, the first federated adaption paradigm for foundation model-based recommendation. It optimizes from a solid starting point, leveraging a pre-trained model as the backbone. Our method involves learning personalized adapters for each client based on local data, focusing on lightweight low-rank adapters at user and user-group levels to learn detailed personalization with complementary perspectives. In addition, we design an adaptive gate learning mechanism to effectively blend common knowledge and user personalization with dynamic weights. Our method significantly reduces computational and communication costs by updating only user-specific parameters during federated optimization. Extensive experiments demonstrate superior performance compared to advanced baselines. We address the challenge of deploying models on resource-constrained devices by distilling a compact model from the original pre-trained model. Additionally, we enhance privacy protection in \baby by incorporating Local Differential Privacy, achieving a solid balance between recommendation performance and privacy preservation. 


\section*{Acknowledgments}
Chunxu Zhang and Bo Yang are supported by the National Key R\&D Program of China under Grant No. 2021ZD0112500; the National Natural Science Foundation of China under Grant Nos. U22A2098, 62172185, 62206105 and 62202200; the Fundamental Research Funds for the Central Universities, JLU; the Key Science and Technology Development Plan of Jilin Province under Grant No. 20240302078GX; Jilin Province Capital Construction Fund Industry Technology Research and Development Project No. 2022C047-1. Yang Liu and Chunxu Zhang are supported by the Tsinghua-Kuaishou Joint Research Program.

\bibliographystyle{named}
\bibliography{ijcai24}

\begin{thebibliography}{}

\bibitem[\protect\citeauthoryear{Achiam \bgroup \em et al.\egroup }{2023}]{OpenAI2023}
Josh Achiam, Steven Adler, Sandhini Agarwal, Lama Ahmad, Ilge Akkaya, Florencia~Leoni Aleman, Diogo Almeida, Janko Altenschmidt, Sam Altman, Shyamal Anadkat, et~al.
\newblock Gpt-4 technical report.
\newblock {\em arXiv preprint arXiv:2303.08774}, 2023.

\bibitem[\protect\citeauthoryear{Aghajanyan \bgroup \em et al.\egroup }{2020}]{aghajanyan2020intrinsic}
Armen Aghajanyan, Luke Zettlemoyer, and Sonal Gupta.
\newblock Intrinsic dimensionality explains the effectiveness of language model fine-tuning.
\newblock {\em arXiv preprint arXiv:2012.13255}, 2020.

\bibitem[\protect\citeauthoryear{Alayrac \bgroup \em et al.\egroup }{2022}]{alayrac2022flamingo}
Jean-Baptiste Alayrac, Jeff Donahue, Pauline Luc, Antoine Miech, Iain Barr, Yana Hasson, Karel Lenc, Arthur Mensch, Katherine Millican, Malcolm Reynolds, et~al.
\newblock Flamingo: a visual language model for few-shot learning.
\newblock {\em Advances in Neural Information Processing Systems}, 35:23716--23736, 2022.

\bibitem[\protect\citeauthoryear{Bommasani \bgroup \em et al.\egroup }{2021}]{bommasani2021opportunities}
Rishi Bommasani, Drew~A Hudson, Ehsan Adeli, Russ Altman, Simran Arora, Sydney von Arx, Michael~S Bernstein, Jeannette Bohg, Antoine Bosselut, Emma Brunskill, et~al.
\newblock On the opportunities and risks of foundation models.
\newblock {\em arXiv preprint arXiv:2108.07258}, 2021.

\bibitem[\protect\citeauthoryear{Brown \bgroup \em et al.\egroup }{2020}]{brown2020language}
Tom Brown, Benjamin Mann, Nick Ryder, Melanie Subbiah, Jared~D Kaplan, Prafulla Dhariwal, Arvind Neelakantan, Pranav Shyam, Girish Sastry, Amanda Askell, et~al.
\newblock Language models are few-shot learners.
\newblock {\em Advances in neural information processing systems}, 33:1877--1901, 2020.

\bibitem[\protect\citeauthoryear{Chai \bgroup \em et al.\egroup }{2020}]{chai2020secure}
Di~Chai, Leye Wang, Kai Chen, and Qiang Yang.
\newblock Secure federated matrix factorization.
\newblock {\em IEEE Intelligent Systems}, 36(5):11--20, 2020.

\bibitem[\protect\citeauthoryear{Choi \bgroup \em et al.\egroup }{2018}]{choi2018guaranteeing}
Woo-Seok Choi, Matthew Tomei, Jose Rodrigo~Sanchez Vicarte, Pavan~Kumar Hanumolu, and Rakesh Kumar.
\newblock Guaranteeing local differential privacy on ultra-low-power systems.
\newblock In {\em 2018 ACM/IEEE 45th Annual International Symposium on Computer Architecture (ISCA)}, pages 561--574. IEEE, 2018.

\bibitem[\protect\citeauthoryear{Devlin \bgroup \em et al.\egroup }{2018}]{devlin2018bert}
Jacob Devlin, Ming-Wei Chang, Kenton Lee, and Kristina Toutanova.
\newblock Bert: Pre-training of deep bidirectional transformers for language understanding.
\newblock {\em arXiv preprint arXiv:1810.04805}, 2018.

\bibitem[\protect\citeauthoryear{Geng \bgroup \em et al.\egroup }{2022}]{geng2022recommendation}
Shijie Geng, Shuchang Liu, Zuohui Fu, Yingqiang Ge, and Yongfeng Zhang.
\newblock Recommendation as language processing (rlp): A unified pretrain, personalized prompt \& predict paradigm (p5).
\newblock In {\em Proceedings of the 16th ACM Conference on Recommender Systems}, pages 299--315, 2022.

\bibitem[\protect\citeauthoryear{Gou \bgroup \em et al.\egroup }{2021}]{gou2021knowledge}
Jianping Gou, Baosheng Yu, Stephen~J Maybank, and Dacheng Tao.
\newblock Knowledge distillation: A survey.
\newblock {\em International Journal of Computer Vision}, 129:1789--1819, 2021.

\bibitem[\protect\citeauthoryear{Harte \bgroup \em et al.\egroup }{2023}]{harte2023leveraging}
Jesse Harte, Wouter Zorgdrager, Panos Louridas, Asterios Katsifodimos, Dietmar Jannach, and Marios Fragkoulis.
\newblock Leveraging large language models for sequential recommendation.
\newblock In {\em Proceedings of the 17th ACM Conference on Recommender Systems}, pages 1096--1102, 2023.

\bibitem[\protect\citeauthoryear{Huang \bgroup \em et al.\egroup }{2023}]{huang2023randomization}
Xinyi Huang, Yuchuan Luo, Lin Liu, Wentao Zhao, and Shaojing Fu.
\newblock Randomization is all you need: A privacy-preserving federated learning framework for news recommendation.
\newblock {\em Information Sciences}, 637:118943, 2023.

\bibitem[\protect\citeauthoryear{Kojima \bgroup \em et al.\egroup }{2022}]{kojima2022large}
Takeshi Kojima, Shixiang~Shane Gu, Machel Reid, Yutaka Matsuo, and Yusuke Iwasawa.
\newblock Large language models are zero-shot reasoners.
\newblock {\em Advances in neural information processing systems}, 35:22199--22213, 2022.

\bibitem[\protect\citeauthoryear{Li \bgroup \em et al.\egroup }{2018}]{li2018measuring}
Chunyuan Li, Heerad Farkhoor, Rosanne Liu, and Jason Yosinski.
\newblock Measuring the intrinsic dimension of objective landscapes.
\newblock In {\em International Conference on Learning Representations}, 2018.

\bibitem[\protect\citeauthoryear{Li \bgroup \em et al.\egroup }{2023}]{li2023federated}
Zhiwei Li, Guodong Long, and Tianyi Zhou.
\newblock Federated recommendation with additive personalization.
\newblock {\em arXiv preprint arXiv:2301.09109}, 2023.

\bibitem[\protect\citeauthoryear{Lin \bgroup \em et al.\egroup }{2020}]{lin2020meta}
Yujie Lin, Pengjie Ren, Zhumin Chen, Zhaochun Ren, Dongxiao Yu, Jun Ma, Maarten~de Rijke, and Xiuzhen Cheng.
\newblock Meta matrix factorization for federated rating predictions.
\newblock In {\em Proceedings of the 43rd International ACM SIGIR Conference on Research and Development in Information Retrieval}, pages 981--990, 2020.

\bibitem[\protect\citeauthoryear{Lin \bgroup \em et al.\egroup }{2023}]{lin2023can}
Jianghao Lin, Xinyi Dai, Yunjia Xi, Weiwen Liu, Bo~Chen, Xiangyang Li, Chenxu Zhu, Huifeng Guo, Yong Yu, Ruiming Tang, et~al.
\newblock How can recommender systems benefit from large language models: A survey.
\newblock {\em arXiv preprint arXiv:2306.05817}, 2023.

\bibitem[\protect\citeauthoryear{Liu \bgroup \em et al.\egroup }{2023a}]{liu2023pre}
Peng Liu, Lemei Zhang, and Jon~Atle Gulla.
\newblock Pre-train, prompt and recommendation: A comprehensive survey of language modelling paradigm adaptations in recommender systems.
\newblock {\em arXiv preprint arXiv:2302.03735}, 2023.

\bibitem[\protect\citeauthoryear{Liu \bgroup \em et al.\egroup }{2023b}]{liu2023privaterec}
Ruixuan Liu, Yang Cao, Yanlin Wang, Lingjuan Lyu, Yun Chen, and Hong Chen.
\newblock Privaterec: Differentially private model training and online serving for federated news recommendation.
\newblock In {\em Proceedings of the 29th ACM SIGKDD Conference on Knowledge Discovery and Data Mining}, pages 4539--4548, 2023.

\bibitem[\protect\citeauthoryear{Liu \bgroup \em et al.\egroup }{2023c}]{liu2023federated}
Weiming Liu, Chaochao Chen, Xinting Liao, Mengling Hu, Jianwei Yin, Yanchao Tan, and Longfei Zheng.
\newblock Federated probabilistic preference distribution modelling with compactness co-clustering for privacy-preserving multi-domain recommendation.
\newblock In {\em Proceedings of the Thirty-Second International Joint Conference on Artificial Intelligence}, pages 2206--2214, 2023.

\bibitem[\protect\citeauthoryear{Liu \bgroup \em et al.\egroup }{2024}]{liu2024icde}
Ziqiao Liu, Hao Miao, Yan Zhao, Chenxi Liu, Kai Zheng, and Huan Li.
\newblock Lighttr: A lightweight framework for federated trajectory recovery.
\newblock In {\em ICDE}, 2024.

\bibitem[\protect\citeauthoryear{McMahan \bgroup \em et al.\egroup }{2017}]{mcmahan2017communication}
Brendan McMahan, Eider Moore, Daniel Ramage, Seth Hampson, and Blaise~Aguera y~Arcas.
\newblock Communication-efficient learning of deep networks from decentralized data.
\newblock In {\em Artificial intelligence and statistics}, pages 1273--1282. PMLR, 2017.

\bibitem[\protect\citeauthoryear{Miao \bgroup \em et al.\egroup }{2023}]{miao2023task}
Hao Miao, Xiaolong Zhong, Jiaxin Liu, Yan Zhao, Xiangyu Zhao, Weizhu Qian, Kai Zheng, and Christian~S Jensen.
\newblock Task assignment with efficient federated preference learning in spatial crowdsourcing.
\newblock {\em IEEE Transactions on Knowledge and Data Engineering}, 2023.

\bibitem[\protect\citeauthoryear{Perifanis and Efraimidis}{2022}]{perifanis2022federated}
Vasileios Perifanis and Pavlos~S Efraimidis.
\newblock Federated neural collaborative filtering.
\newblock {\em Knowledge-Based Systems}, 242:108441, 2022.

\bibitem[\protect\citeauthoryear{Pillutla \bgroup \em et al.\egroup }{2022}]{pillutla2022federated}
Krishna Pillutla, Kshitiz Malik, Abdel-Rahman Mohamed, Mike Rabbat, Maziar Sanjabi, and Lin Xiao.
\newblock Federated learning with partial model personalization.
\newblock In {\em International Conference on Machine Learning}, pages 17716--17758. PMLR, 2022.

\bibitem[\protect\citeauthoryear{Qiu \bgroup \em et al.\egroup }{2020}]{qiu2020pre}
Xipeng Qiu, Tianxiang Sun, Yige Xu, Yunfan Shao, Ning Dai, and Xuanjing Huang.
\newblock Pre-trained models for natural language processing: A survey.
\newblock {\em Science China Technological Sciences}, 63(10):1872--1897, 2020.

\bibitem[\protect\citeauthoryear{Qu \bgroup \em et al.\egroup }{2023}]{qu2023semi}
Liang Qu, Ningzhi Tang, Ruiqi Zheng, Quoc Viet~Hung Nguyen, Zi~Huang, Yuhui Shi, and Hongzhi Yin.
\newblock Semi-decentralized federated ego graph learning for recommendation.
\newblock In {\em Proceedings of the ACM Web Conference}, page 339–348, 2023.

\bibitem[\protect\citeauthoryear{Radford \bgroup \em et al.\egroup }{2019}]{radford2019language}
Alec Radford, Jeffrey Wu, Rewon Child, David Luan, Dario Amodei, Ilya Sutskever, et~al.
\newblock Language models are unsupervised multitask learners.
\newblock {\em OpenAI blog}, 1(8):9, 2019.

\bibitem[\protect\citeauthoryear{Saharia \bgroup \em et al.\egroup }{2022}]{saharia2022photorealistic}
Chitwan Saharia, William Chan, Saurabh Saxena, Lala Li, Jay Whang, Emily~L Denton, Kamyar Ghasemipour, Raphael Gontijo~Lopes, Burcu Karagol~Ayan, Tim Salimans, et~al.
\newblock Photorealistic text-to-image diffusion models with deep language understanding.
\newblock {\em Advances in Neural Information Processing Systems}, 35:36479--36494, 2022.

\bibitem[\protect\citeauthoryear{Singhal \bgroup \em et al.\egroup }{2021}]{singhal2021federated}
Karan Singhal, Hakim Sidahmed, Zachary Garrett, Shanshan Wu, John Rush, and Sushant Prakash.
\newblock Federated reconstruction: Partially local federated learning.
\newblock {\em Advances in Neural Information Processing Systems}, 34:11220--11232, 2021.

\bibitem[\protect\citeauthoryear{Sun \bgroup \em et al.\egroup }{2022}]{sun2022survey}
Zehua Sun, Yonghui Xu, Yong Liu, Wei He, Lanju Kong, Fangzhao Wu, Yali Jiang, and Lizhen Cui.
\newblock A survey on federated recommendation systems.
\newblock {\em arXiv preprint arXiv:2301.00767}, 2022.

\bibitem[\protect\citeauthoryear{Wang \bgroup \em et al.\egroup }{2020}]{wang2020tackling}
Jianyu Wang, Qinghua Liu, Hao Liang, Gauri Joshi, and H~Vincent Poor.
\newblock Tackling the objective inconsistency problem in heterogeneous federated optimization.
\newblock {\em Advances in neural information processing systems}, 33:7611--7623, 2020.

\bibitem[\protect\citeauthoryear{Wu \bgroup \em et al.\egroup }{2022}]{wu2022federated}
Chuhan Wu, Fangzhao Wu, Lingjuan Lyu, Tao Qi, Yongfeng Huang, and Xing Xie.
\newblock A federated graph neural network framework for privacy-preserving personalization.
\newblock {\em Nature Communications}, 13(1):3091, 2022.

\bibitem[\protect\citeauthoryear{Wu \bgroup \em et al.\egroup }{2023}]{wu2023survey}
Likang Wu, Zhi Zheng, Zhaopeng Qiu, Hao Wang, Hongchao Gu, Tingjia Shen, Chuan Qin, Chen Zhu, Hengshu Zhu, Qi~Liu, et~al.
\newblock A survey on large language models for recommendation.
\newblock {\em arXiv preprint arXiv:2305.19860}, 2023.

\bibitem[\protect\citeauthoryear{Yang \bgroup \em et al.\egroup }{2020}]{yang2020federated}
Liu Yang, Ben Tan, Vincent~W Zheng, Kai Chen, and Qiang Yang.
\newblock Federated recommendation systems.
\newblock {\em Federated Learning: Privacy and Incentive}, pages 225--239, 2020.

\bibitem[\protect\citeauthoryear{Zhang \bgroup \em et al.\egroup }{2023a}]{zhang2023dual}
Chunxu Zhang, Guodong Long, Tianyi Zhou, Peng Yan, Zijian Zhang, Chengqi Zhang, and Bo~Yang.
\newblock Dual personalization on federated recommendation.
\newblock {\em arXiv preprint arXiv:2301.08143}, 2023.

\bibitem[\protect\citeauthoryear{Zhang \bgroup \em et al.\egroup }{2023b}]{zhang2023ifedrec}
Chunxu Zhang, Guodong Long, Tianyi Zhou, Xiangyu Zhao, Zijian Zhang, and Bo~Yang.
\newblock When federated recommendation meets cold-start problem: Separating item attributes and user interactions.
\newblock {\em arXiv preprint arXiv:2305.12650}, 2023.

\bibitem[\protect\citeauthoryear{Zhang \bgroup \em et al.\egroup }{2023c}]{zhang2023fine}
Xiao Zhang, Ziming Ye, Jianfeng Lu, Fuzhen Zhuang, Yanwei Zheng, and Dongxiao Yu.
\newblock Fine-grained preference-aware personalized federated poi recommendation with data sparsity.
\newblock In {\em Proceedings of the 46th International ACM SIGIR Conference on Research and Development in Information Retrieval}, pages 413--422, 2023.

\bibitem[\protect\citeauthoryear{Zhong \bgroup \em et al.\egroup }{2023}]{zhong2023personalized}
Xiaolong Zhong, Hao Miao, Dazhuo Qiu, Yan Zhao, and Kai Zheng.
\newblock Personalized location-preference learning for federated task assignment in spatial crowdsourcing.
\newblock In {\em Proceedings of the 32nd ACM International Conference on Information and Knowledge Management}, pages 3534--3543, 2023.

\end{thebibliography}

\end{document}